\documentclass[reprint,
superscriptaddress,
amsmath,amssymb,
 aip,
]{revtex4-2}
\usepackage{xcolor}
\usepackage{amsmath}
\usepackage{mathptmx} 
\usepackage{xfrac}
\usepackage{gensymb}
\usepackage{upgreek}
\usepackage{graphicx}
\usepackage{dcolumn}
\usepackage{bm}
\usepackage{hyperref}

\begin{document}
\title{Photophysics of O-band and transition metal color centers in monolithic silicon for quantum communications}
\author{Murat Can Sarihan}
\email{mcansarihan@gmail.com}
\thanks{These authors contributed equally.}
\affiliation{Department of Electrical and Computer Engineering, University of California, Los Angeles, California, USA}%
\author{Jiahui Huang}
\thanks{These authors contributed equally.}
\affiliation{Department of Electrical and Computer Engineering, University of California, Los Angeles, California, USA}%

\author{Jin Ho Kang}
\affiliation{Department of Electrical and Computer Engineering, University of California, Los Angeles, California, USA}%
\author{Cody Fan}
\affiliation{Department of Electrical and Computer Engineering, University of California, Los Angeles, California, USA}%
\author{Wei Liu}
\affiliation{Department of Electrical and Computer Engineering, University of California, Los Angeles, California, USA}%
\author{Khalifa M. Azizur-Rahman}
\affiliation{Department of Electrical and Computer Engineering, University of California, Los Angeles, California, USA}%
\affiliation{Center for Integrated Nanotechnologies, Sandia National Laboratories, Albuquerque, New Mexico, USA}
\author{Baolai Liang}
\affiliation{Department of Electrical and Computer Engineering, University of California, Los Angeles, California, USA}%
\author{Chee Wei Wong}
\affiliation{Department of Electrical and Computer Engineering, University of California, Los Angeles, California, USA}%

\begin{abstract}
Color centers in the O-band (1260-1360 nm) are crucial for realizing long-coherence quantum network nodes in memory-assisted quantum communications. However, only a limited number of O-band color centers have been thoroughly explored in silicon hosts as spin-photon interfaces. This study explores and compares two promising O-band color centers in silicon for high-fidelity spin-photon interfaces: T and $^*$Cu (transition metal) centers. During T center generation process, we observed the formation and dissolution of other color centers, including the copper-silver related centers with a doublet line around 1312 nm ($^*$Cu$^{0}_{n}$), near the optical fiber zero dispersion wavelength (around 1310 nm). We then investigated the photophysics of both T and $^*$Cu centers, focusing on their emission spectra and spin properties. The $^*$Cu$^{0}_{0}$ line under a 0.5 T magnetic field demonstrated a 25\% broadening, potentially due to spin degeneracy, suggesting that this center can be a promising alternative to T centers.

\end{abstract}
\maketitle
\section{Introduction}
Building a practical and efficient quantum network is essential to scaling the computational power of quantum computers, leading to the wider application and development of secure quantum communications \cite{zhouLPR2022,Rempe.2012.Nature,RempeRMP2015,AwschalomNRM2021}. To efficiently send flying qubits (photons) over long distances in an optical fiber network, their particular wavelengths need to be either centered around the lowest dispersion wavelength of 1310 nm (O-band) or the lowest loss wavelength of 1550 nm (C-band). It is also important to state that an intermediate node such as a quantum repeater is necessary to optimize the flying qubits' communication and network range \cite{Hanson.2022.Nature, nemotoAVS2022,Reisererarxiv2022,lukinNature2020, noriarxiv2023,senellartarxiv2022,lukinPRL2022,Lukin.2022c9o,Hosseini.2023}. For optimization, it is considered that high dimensional energy-time entangled qubits is a promising platform for carrying quantum information efficiently with high information capacity between high fidelity quantum memory nodes \cite{Oxenløwe.201929l,Zeilinger.2020,Wong.2021o492,Dolecek.2019}. Additionally, it is also considered that for low error rate communications, operating the high dimensional quantum channels near 1310 nm is desirable due to minimal timing errors over long distances, increasing state fidelity. It is essential to develop a high fidelity quantum node with near unity radiative efficiency devoid of possible nonradiative pathways that can be integrated with on-chip photonic devices to interface with O-band quantum channels efficiently. Further to this discussion, the radiative lifetime of the defect centers also must be decreased by integrating them into cavity structures in order to achieve high coherence fast photonic qubit interactions.

Solid state color center qubits in silicon have been examined as prominent candidates for quantum nodes for several reasons \cite{hautierarxiv2023,Rogge.2023,
Reisererarxiv2023}. Silicon is the dominant, scalable, and well researched platform for nanoelectronic and nanophotonic circuits to control and interface with qubits. It is possible to create reproducible color centers and donor defects in silicon with exceptional characteristics. Furthermore, recently developed $^{28}$Si substrates that are isotopically purified up to $99.9998\%$ offer a host devoid of magnetic field fluctuations, supporting good qubit coherence \cite{pomeroyAIP2019,pomeroyJPC2020,pomeroyJPD2014,pomeroyPRM2017,pomeroyRSI2019}.  In silicon hosts, spin qubits with long lifetimes of up to 3 hours for shallow phosphorus donors \cite{thewaltScience2013} and 2.14 seconds for deep selenium donors \cite{SimmonsSciAdv2017, SimmonsPRApp2019} have been developed. Phosphorus donors operate in the microwave region, and selenium donors have a mid-infrared transition, both mechanisms of which are less expeditious to interface with current fiber network infrastructure. 

Color centers in silicon, such as G, W, and T centers, are promising due to their easily reproducible defect generation processes, bright photoluminescence (PL) emissions near the O-band, long coherence time, narrow emission linewidths, and their ability to be integrated into silicon photonic devices \cite{redjemarxiv2023,CassaboisPRB2018,CassaboisNatElec2020,DreauPRL2021,leeArxiv2021,hautierarxiv2023,liuPRApp2023,dreauACS2022,astakhovNComm2022,englundarxiv2023,
dreau2023_APL}. G and W centers have bright single photon emission owing to their singlet-singlet transitions \cite{Thewalt.2018.Physical,englundarxiv2023,
dreau2023_APL}. These defects possess great potential as highly indistinguishable single photon sources for integrated quantum photonics. G centers in particular also exhibit a metastable triplet excited state transition as a possible spin photon interface \cite{PhysRevLett.48.37}.

Alternatively, T centers have shown promising properties that make them suitable as efficient and coherent spin photon interfaces in the telecommunication wavelengths \cite{simmons2020,simmonsarXiv2022_wg, simmonsarXiv2022_memory,Wong.2022,WongCLEO2023,Wong.2021nsl,cavity1,cavity2,cavity3,simmons2024_1,simmons2024_2}. T centers have a doublet line (TX$_{0}$ and TX$_{1}$) with 1.76 meV spacing, most likely resulting from internal stress \cite{safonovPRL1996}. T centers consist of two carbon atoms occupying a single silicon site, while a hydrogen atom is bonded with one of the carbon atoms forming a $C_{1h}$ symmetry \cite{song1990bistable}. The T center is hypothesized to be formed by capturing an interstitial carbon-hydrogen complex at a substitutional carbon site \cite{obergPRB1998}. Electronically, the T center site includes a bound exciton and an unpaired electron, where two electrons constitute a spin-0 singlet state, while the unpaired hole has a spin-3/2 state \cite{safonovPRL1996}.

The first (TX$_{0}$) and second (TX$_{1}$) excited state transitions of T centers were reported to be at $\approx$ 1325.9 nm (935.1 meV) and 1323.4 nm (936.9 meV) respectively, close to the O-band zero dispersion wavelength. The ensemble linewidth of TX$_{0}$ zero-phonon line (ZPL) of T centers was reported as 26.9 $\mathrm{\upmu}$eV in natural silicon and 0.14 $\mathrm{\upmu}$eV in $^{28}$Si \cite{simmons2020}. The reported TX$_{0}$ lifetime was 0.94 $\mathrm{\upmu}$s, leading to a possible ZPL dipole moment of 0.73 Debye and a Debye-Waller factor of 0.23 at 1.4 K \cite{simmons2020}. Spin resonance measurements have shown a nuclear spin coherence time of over a second from Hahn echo measurements \cite{simmons2020}. The aforementioned properties make T centers an attractive platform for a metropolitan scale quantum network.

A less studied possible candidate for spin-photon interfaces is a Cu-Ag transition metal defect in silicon which is known as the $^*$Cu defect. The $^*$Cu defect has been shown to have a doublet emission at approximately 1312.15 nm, which is closer to the zero dispersion wavelength compared to T centers \cite{StegerJAP2011, Thewalt.1988, Lightowlers.1989}. This was first discovered by McGuigan et al. in a lightly Cu doped silicon substrate and reported to exhibit an isoelectronic bound exciton nature in the ground states \cite{StegerJAP2011, Thewalt.1988, Lightowlers.1989}. However, the defect induced lattice distortion, as well as the photophysics and radiative dynamics of its bound excitons, has not been as well studied as T centers.

In this work, we examined and identified the relevant transformations between the plethora of color centers in the process of generating Cu related defects and T centers in silicon. Our sequential observation of Cu related defects and T centers, during fabrication on the same silicon substrate, made comparison possible between the two color centers. We then studied the photophysics of the  $^*$Cu doublet transitions, which appeared in the intermediate process steps, using cryogenic microphotoluminescence ($\mathrm{\upmu}$PL) spectroscopy and we compared it with the subsequently generated T centers on the same substrate. Our analysis of the temperature dependent PL suggested that of the two peaks of the $^*$Cu doublet transitions, one was associated with at least one strongly localized state and the other with a loosely bound state. Electron spin resonance and magneto-PL measurements on the $^*$Cu transitions revealed an effective Zeeman factor of 364 $\mathrm{\upmu}$eV/T and an ensemble g-factor of 2.002, which demonstrates the potential of using $^*$Cu defects as a potential alternative candidate to T centers for a spin-photon interface.

\begin{figure*}
    \includegraphics[width=\textwidth]{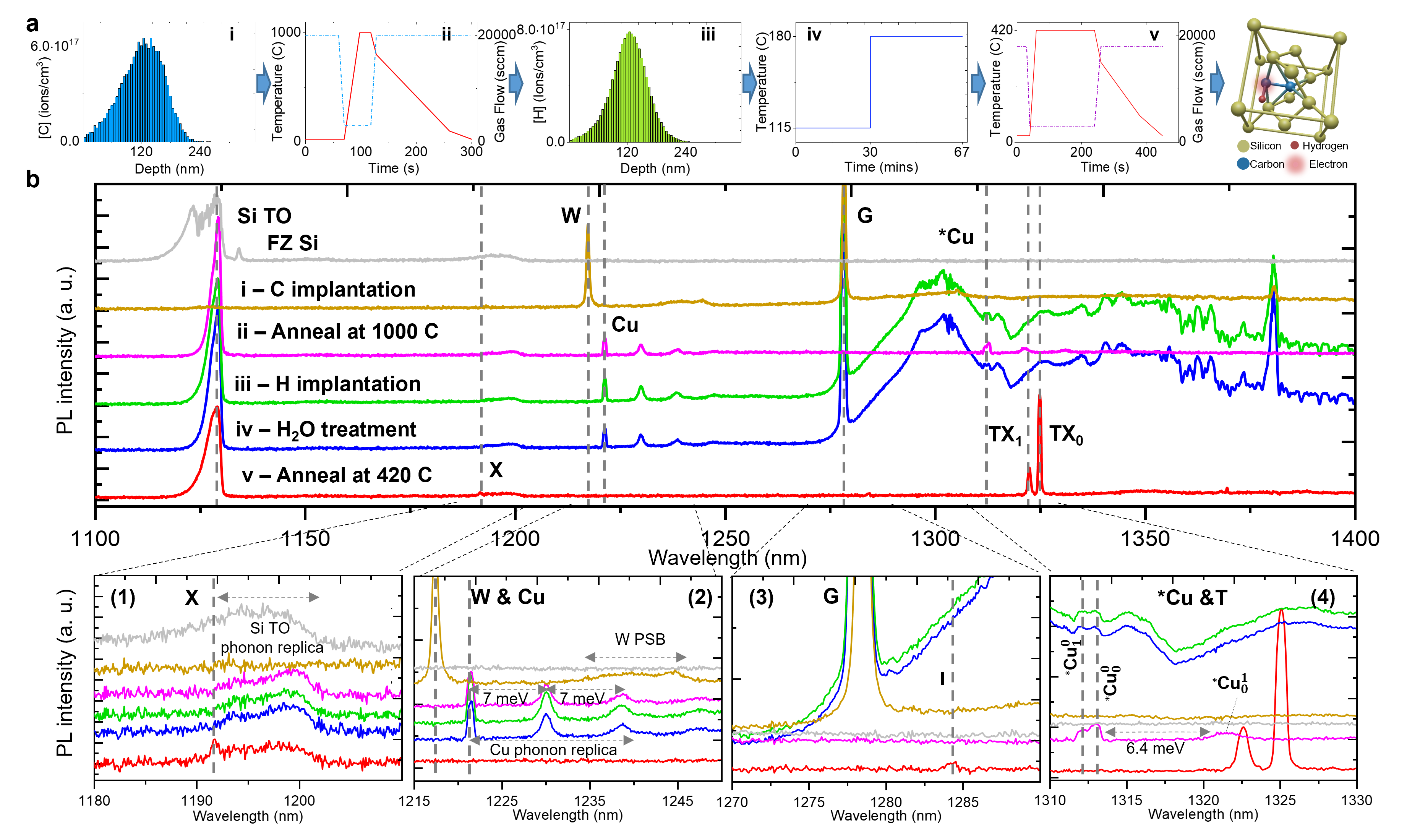}
    \caption{\textbf{Family of color centers in silicon and the T center generation recipe.}  \textbf{(a)} Process steps: i) Simulated depth profile of carbon implantation with 38 keV ion energy. ii) Temperature (red line) and gas flow (blue dashed line) profiles for rapid thermal annealing recipe at 1000 \degree C for 20 seconds. iii) Simulated depth profile of hydrogen implantation with 9 keV ion energy. iv) Hot plate temperature variation for H$_2$O treatment. v) Temperature (red line) and gas flow (blue dashed line) profiles for rapid thermal annealing at 420 \degree C for 3 minutes. \textbf{(b)} Change in the microphotoluminescence ($\mathrm{\upmu}$PL) spectrum at each process step, where each identified line is marked with vertical dashed lines and the spectra are stacked with equal spacing vertically for easier identification. The X, W, G, and T emission lines are magnified at the bottom.  The observed 0.56 meV split $^*$Cu doublet at approximately 1312.15 nm is denoted as the low ($^*$Cu$_0^0$) and high ($^*$Cu$_1^0$) energy line. All spectra were taken at 5 K using a 532 nm continuous wave (CW) laser with 900 W/cm$^2$ intensity and 600 grooves/mm grating.}
    \label{fig:fig1}
\end{figure*}

\section{Results and Discussion}
\subsection{Fabrication process and transformation of color centers}

We examined a combination of ion implantation and rapid thermal annealing processes to control and optimize the generation of Cu-related defects and T centers. Our process was adapted from earlier studies \cite{simmonsArxiv2021, simmonsNJP2021}. A float-zone (FZ) grown, dilute boron-doped silicon substrate was used, with a resistivity of 1000 $\Omega\cdot \textrm{cm}$ to 2000 $\Omega\cdot \textrm{cm}$ with carbon and hydrogen impurity concentrations less than $2 \times 10^{16}$ cm$^{-3}$. First, carbon ions were implanted into a silicon substrate at 7\degree{} with 38 keV energy and a dose of $7 \times 10^{12}$ cm$^{-2}$. This implantation recipe produced an ion profile whose peak concentration was at 118.8 nm from the surface with a 42.8 nm straggle based on SRIM-2013 simulations, with the profile shown in Figure \ref{fig:fig1}a-i \cite{ZIEGLER20101818}.

The lattice damage from carbon implantation was repaired by rapid thermal annealing (RTA) at 1000 \degree C for 20 seconds in an argon environment with a flow rate of 3000 sccm. The maximum temperature was achieved after a 28 second ramp up, while the cooling was done passively under high argon flow (20000 sccm). Figure \ref{fig:fig1}a-ii shows the temperature profile used.  This recipe is commonly used to introduce substitutional carbon centers in silicon to form G centers\cite{CassaboisNatElec2020,CassaboisPRB2018,Thewalt.2018.Physical}. In order to promote the C-H groups needed to form the T centers, we implanted hydrogen ions into the silicon substrate at 7$\degree$ with 9 keV energy and a dose of $7 \times 10^{12}$ cm$^{-2}$. As a result of this implantation, we obtained a hydrogen concentration profile illustrated in Figure \ref{fig:fig1}a-iii with a peak concentration at 120.8 nm from the surface and with a straggle of 38 nm, matching the carbon concentration \cite{ZIEGLER20101818}. Thereafter, the samples were cleaned using acetone and isopropyl alcohol then N$_2$ blow dried.

Afterward, the samples were treated in deionized H$_2$O to promote hydrogen diffusion and bonding with carbon atoms. The treatment was done in a cleanroom setting under a fume hood, using a hot plate with a two step process. The samples were treated in deionized H$_2$O with a hot plate temperature of 115 \degree C for 30 minutes, and subsequently with the hot plate increased to 180 \degree C and held for 37 minutes (Figure \ref{fig:fig1}a-iv). This two step treatment was performed under the fume hood and was designed to apply a controllable amount of thermal energy, with H$_2$O surface temperatures up to 95.5 \degree C. The treated samples were then placed into the RTA chamber for the last step to promote the binding of C-H groups with substitutional carbon sites to form T centers. Inside the nitrogen environment, the samples were annealed at 420 \degree C for 3 minutes within a SiC susceptor.

Figure \ref{fig:fig1}b shows the $\mathrm{\upmu}$PL spectrum taken between 1100 nm and 1400 nm at 5 K after each process step to examine the Cu-related defect and T center formation mechanism and their precursors compared to untreated Si substrate.  As a benchmark of lattice damage in silicon, it was observed that the silicon transverse-optical (TO) transition line around 1130 nm (bandgap) was completely destroyed upon carbon implantation, which was recovered by RTA in step ii \cite{henryJAP1991}. Owing to the low atomic mass, hydrogen implantation only showed a 10\% decrease in the Si TO line compared to step ii. The deionized H$_2$O treatment also acted as an annealing process, which enhanced the Si TO line damaged by hydrogen implantation. After the second RTA step, we observed a 30\% decrease in Si TO line intensity, which might have been due to replacement of the Si sites with the C-(C-H) complex that led to the T center formation \cite{safonovPRL1996}. We observed a similar trend in the Si phonon replica around 1195 nm. Superimposed on the phonon replica, after step v, we observed the weak X center line at 1191.7 nm. The origins of the X center were uncertain; initial studies had suggested an interstitial nature with a possible tetrahedral or tetragonal symmetry, while the recent ab initio studies on candidates such as I3-X or I4-A defects were inconclusive \cite{santosJPD2016}.

After carbon implantation at step i, before rapid annealing of the resulting lattice damage, we clearly observed the formation of W and G centers whose ZPL transitions were at 1217.48 nm and 1278.5 nm, respectively, with Gaussian full width at half maximum (FWHM) linewidths at approximately 0.65 nm and 0.8 nm. The G center is a carbon related defect, thought to be composed of two substitutional carbon atoms connected by an interstitial silicon atom \cite{CassaboisNatElec2020, CassaboisPRB2018}. Here, we observed the G line clearly without any annealing. The lifetime of the produced G center was also measured up to 6.67 ns at 7.5 K (Figure S7 in Supplementary Information), which was slightly longer than the 5.9 ns previously reported for the ensembles but indicating the dominance of nonradiative processes\cite{CassaboisPRB2018, liuPRApp2023}. The W line had a phonon sideband up to 1250 nm, while the G line phonon sideband extended up to 1400 nm, including a local vibrational mode at 1381 nm \cite{ivanov2022effect}. Annealing at 1000 \degree C for 20 seconds, as shown in \ref{fig:fig1}a-ii, however, led to the disappearance of the G and W lines. This disappearance indicated the displacement of implantation induced silicon interstitials into the lattice during thermal annealing. The W line is thought to be a cluster of interstitial silicon atoms in the lattice with a ZPL at approximately 1218 nm \cite{dreauACS2022,cherkova2019luminescence,davies1987Wline,santosJPD2016}.

The W line was replaced in step ii by three phonon replicas at 1221.45 nm, 1230.07 nm, and 1238.64 nm with FWHM of 0.58 nm, 1.2 nm, and 1.65 nm, respectively, as shown in Figure 1b. These lines were approximately 7 meV apart, indicating the energy of the involved phonons. The peak energies and separation are aligned with prior literature where it is identified as the Cu line, a complex of four Cu atoms \cite{Thewalt.1988, Carvalho.2012, Shirai.2021, StegerJAP2011}. We observed a similar structure for the $^*$Cu$^{m}_{n}$ doublet peak around 1312.15 nm (Figure \ref{fig:fig1}b-(4), consisting of a 0.56 meV split into low ($^*$Cu$_0^0$) and high ($^*$Cu$_1^0$) energy lines at 1312.98 nm and 1312.20 nm respectively. Here, the superscript, m, indicates the number of phonons involved in the transition, hence denoting the phonon replicas, while the subscript, n, indicates the order of the transition in the excitonic energy ladder. We concurrently observed 6.4 meV phonon replicas at 1321.64 nm (called $^*$Cu$^1_0$) and 1331.06 nm (called $^*$Cu$_0^2$). The full list of zero-phonon and phonon-assisted transitions is given in Section L of Supplementary Information. The observed spectral peaks were formed after the carbon implantation and RTA at 1000 \degree{}C, after step ii. The depth profiles were collected using secondary ion mass spectroscopy (SIMS) for steps i and ii, and are shown in Figure S8 in Supplementary Information. Cu and Ag residues were observed, and we suspect their presence was due to the manufacturing process. We speculate that the formation of Cu and $^*$Cu$^{m}_{n}$ was due to the atomic rearrangement during our annealing process. The Gaussian FWHM linewidth of the $^*$Cu$_1^0$ peak was approximately 0.66 nm, while the $^*$Cu$_0^0$ peak had a linewidth of 0.523 nm. These peaks disappeared when deionized H$_2$O treatment was applied.

The G center emission was recovered after hydrogen implantation. It resulted from the implanted proton induced interstitials and persisted after the deionized H$_2$O treatment \cite{qubs6010013}. However, it had a high intensity phonon sideband from 1280 nm to 1400 nm, with the previously reported E-line evident at 1381 nm \cite{ivanov2022effect,thonke1981new}, as observed in step i. Furthermore, after step v, a small peak was observed around 1285 nm, corresponding to the I-center, a variation of the T center perturbed by the presence of oxygen atoms \cite{gower1997centre}. The oxygen concentration $[O]$ was less than  $2 \times 10^{16}$ cm$^{-3}$ in the substrate. As a result of the two step H$_2$O treatment and RTA at 420 °C for 3 minutes, we obtained a clean TX$_0$ zero-phonon line at 1325.1 nm with 0.63 nm linewidth and a TX$_1$ line at 1322.65 nm with 0.79 nm linewidth.

\subsection{Pump power dependent microphotoluminescence spectroscopy of T center and $^*$Cu defects}
\begin{figure*}[]
    \includegraphics[width=\textwidth]{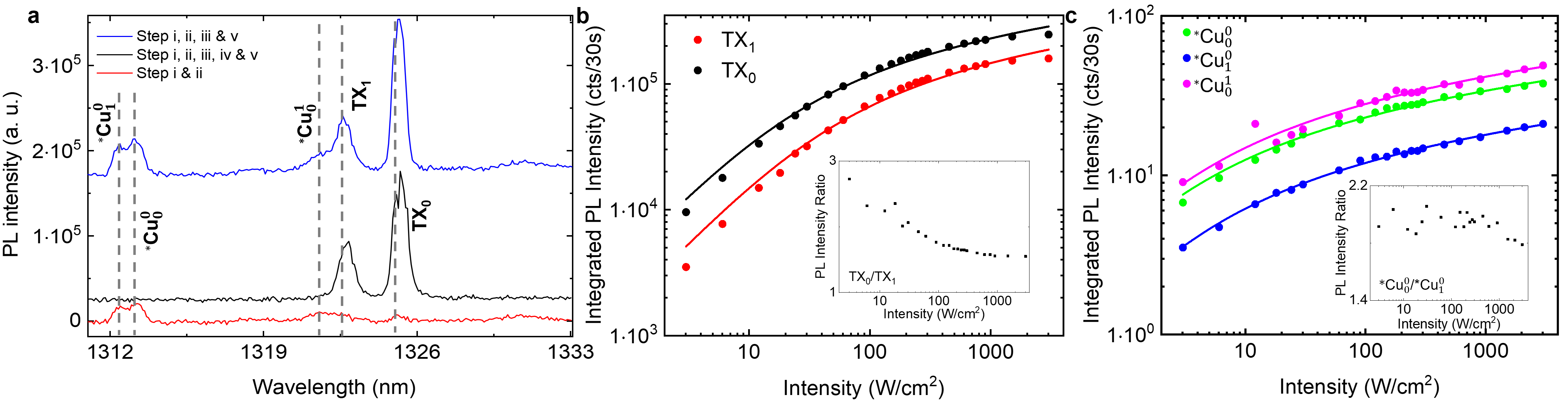}
    \caption{\textbf{Cryogenic microphotoluminescence ($\mathrm{\upmu}$PL) characteristics of TX$_0$, TX$_1$ and $^*$Cu$^{m}_{n}$ transitions in silicon.} \textbf{(a)} PL spectrum showing the formation of TX$_0$, TX$_1$, $^*$Cu$_0^0$, $^*$Cu$_1^0$, and $^*$Cu$_0^1$ peaks in samples processed with different process steps. \textbf{(b)} Pump power dependence of the TX$_0$ (black line) and TX$_1$ (red line) PL integrated intensities. The solid lines show the fit of integrated intensity for each line. Inset: ratio of TX$_0$/TX$_1$ intensities with increasing pump power. \textbf{(c)} Pump power dependence of the $^*$Cu$_0^0$ (green line), $^*$Cu$_1^0$ (blue line) and $^*$Cu$_0^1$ (purple line) PL integrated intensities. Inset: ratio of $^*$Cu$_0^0$/$^*$Cu$_1^0$ intensity with increasing pump power.
}
    \label{fig:fig2}
\end{figure*}

 Figure \ref{fig:fig2}a shows the TX$_{0}$ and TX$_{1}$ lines in the samples that were carbon and hydrogen implanted and H$_2$O and RTA treated (black curve), and also the case where the samples were RTA treated only after step iii (without H$_2$O treatment, blue curve). T center peaks were still observed without the H$_2$O treatment. This indicates that, although deionized H$_2$O treatment facilitates high-purity sample development, it is not required to induce T center formation\cite{simmonsNJP2021}. When we applied step v RTA process, without H$_2$O treatment, we were able to observe  $^*$Cu$_0^0$ and $^*$Cu$_1^0$ peaks alongside the T center TX$_0$ and TX$_1$ lines, while the phonon replica $^*$Cu$_0^1$ at 1321.64 nm showed itself as a shoulder peak to TX$_1$. Hence we speculated that the $^*$Cu centers were susceptible to hydrogenation either via implantation or H$_2$O treatment used for hydrogen diffusion. We concluded that the H$_2$O treatment (step iv) can eliminate the $^*$Cu emission while preserving the T centers.

The integrated intensity of TX$_{0}$ and TX$_{1}$ emissions as a function of pump power is given in Figure \ref{fig:fig2}b. The saturation power was calculated from the fits as 7.8 W/cm$^2$  and 14.3 W/cm$^2$ for TX$_{0}$ and TX$_{1}$ respectively \cite{CassaboisPRB2018}. Presented in Figure \ref{fig:fig2}b inset, at 30 W/cm$^2$ pump power, the ratio of ZPL intensity between TX$_{0}$ and TX$_{1}$ started at approximately 2.725, which gradually decreased and reached a minimum plateau point at approximately 900 W/cm$^2$ pump power where TX$_0$ states saturated. The optically injected carriers kept populating the TX$_1$ states up to 1500 W/cm$^2$.

The same analysis was done for $^*$Cu$_0^0$ and $^*$Cu$_1^0$ peaks and the shoulder $^*$Cu$_0^1$ peak. Like TX$_{0}$ and TX$_{1}$, the $^*$Cu$_0^0$ and $^*$Cu$_1^0$ intensities exhibited saturation at high pump power. Note that the broad $^*$Cu$_0^1$ line, at the lower energy side of the $^*$Cu$_0^0$ and $^*$Cu$_1^0$ lines, followed the same trend as the $^*$Cu$_0^0$ and $^*$Cu$_1^0$ lines, suggesting that the $^*$Cu$_0^1$ line can be attributed to a phonon replica. Unlike the T center, the intensity ratio between $^*$Cu$_0^0$ and $^*$Cu$_1^0$ began at approximately 2, and stayed constant up to 300 W/cm$^2$, and then gradually decreased at higher power. Note that Cu center phonon replicas appeared at the higher energy side of $^*$Cu$_0^0$ and $^*$Cu$_1^0$ lines between 1221 nm and 1240 nm (Figure \ref{fig:fig1}b-2) while there was no visible T center spectrum (Figure \ref{fig:fig1}b-4) during process steps ii to iv. 

\subsection{Temperature dependent microphotoluminescence spectroscopy on T center and $^*$Cu defects}

\begin{figure*}
    \includegraphics[width=\textwidth]{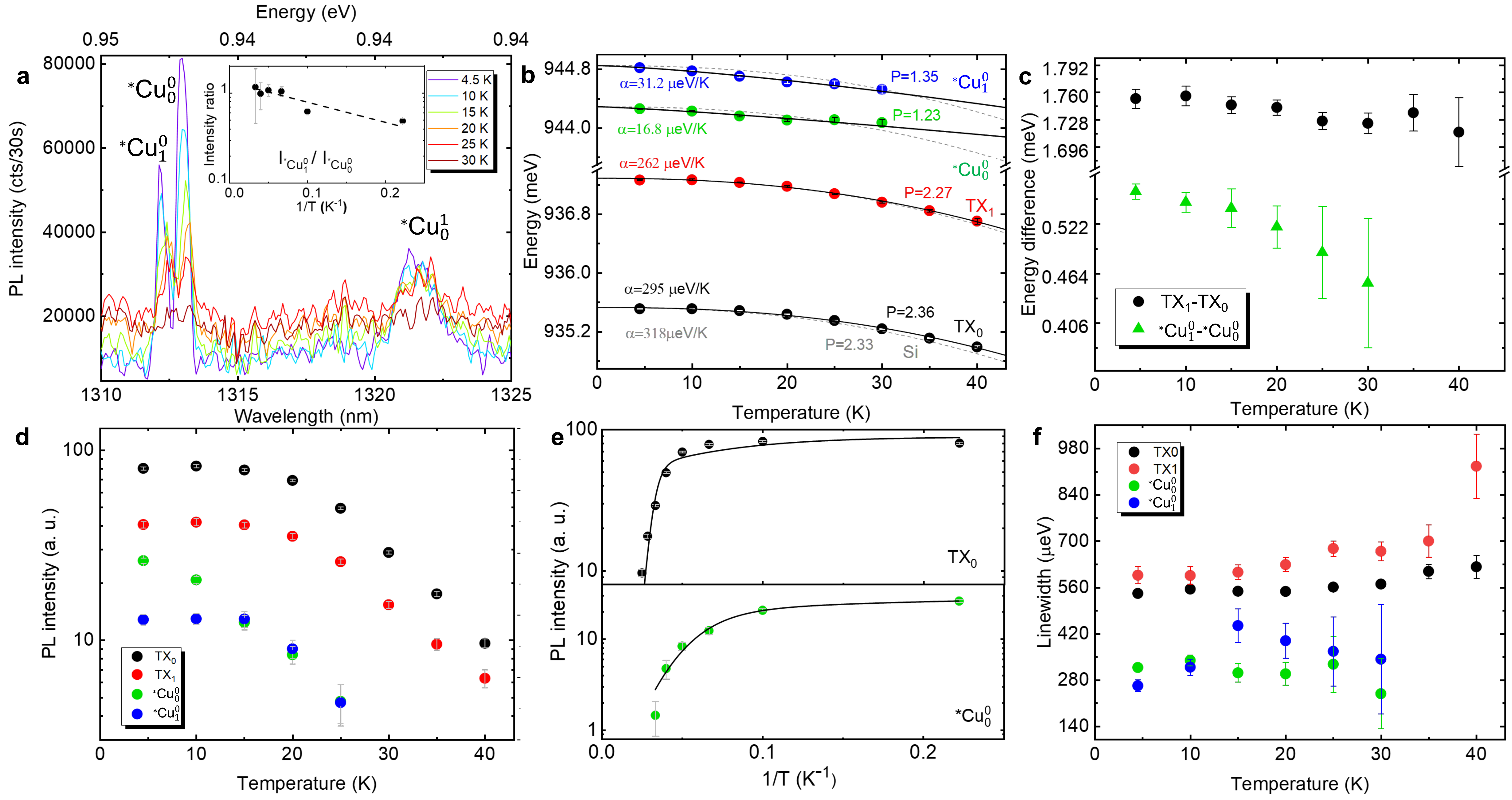}
    \caption{\textbf{Temperature dependence of microphotoluminescence ($\mathrm{\upmu}$PL) on T center zero-phonon line (ZPL) (TX$_{0}$ and TX$_{1}$) and $^*$Cu doublet at 1.313 $\mathrm{\upmu}$m ($^*$Cu$_0^0$ and $^*$Cu$_1^0$).} \textbf{(a)}  Temperature dependent $\mathrm{\upmu}$PL spectra of $^*$Cu$_0^1$, $^*$Cu$_0^0$ and $^*$Cu$_1^0$ lines. Inset: Intensity ratio of $^*$Cu$_1^0$/$^*$Cu$_0^0$ as a function of reciprocal temperature. \textbf{(b)}  ZPL transition energy of TX$_{0}$ (black dots), TX$_{1}$ (red dots), $^*$Cu$_0^0$ (green dots), and $^*$Cu$_1^0$ (blue dots) as a function of temperature. The four black solid curves denote the fit by the four parameter Passler model \cite{passler1999parameter}. The grey dashed curves are for bulk silicon. \textbf{(c)}  The energy difference between $^*$Cu$_0^0$ to  $^*$Cu$_0^1$ (green dots) and TX$_{0}$ to TX$_{1}$ (black dots) as a function of temperature. \textbf{(d)}  TX$_{0}$ (black dots), TX$_{1}$ (red dots), $^*$Cu$_0^0$ (green dots), and $^*$Cu$_1^0$ (blue dots) integrated ZPL intensity as a function of temperature.  \textbf{(e)}  The integrated PL intensity of TX$_{0}$ and $^*$Cu$_0^0$ as a function of reciprocal temperature. The solid black curves denote the fits by Eq. (3). \textbf{(f)}  TX$_{0}$ (black dots), TX$_{1}$ (red dots), $^*$Cu$_0^0$ (green dots), and $^*$Cu$_1^0$ (blue dots) linewidth as a function of temperature. The error bars in the figure denote the standard deviation of the fitting parameter. Excitation laser: 532 nm, power: 900 W/cm$^2$.}
    \label{fig:fig3}
\end{figure*}

To investigate the nature of the observed $^*$Cu doublet emission at 1312.15 nm, we performed a comparative study of the temperature dependent $\mathrm{\upmu}$PL spectroscopy measurement on the T center ZPL. Under low injection below $\mathrm{\upmu}$PL intensity saturation, Figure \ref{fig:fig3}a shows the temperature dependent $\mathrm{\upmu}$PL spectra of the $^*$Cu$_0^1$, $^*$Cu$_0^0$ and $^*$Cu$_1^0$ lines from 4.5 K to 30 K on an FZ silicon sample with carbon implantation and subsequent RTA process detailed in steps i and ii of Figure \ref{fig:fig1}a. A similar temperature dependent measurement of the T center ZPL (TX$_{0}$ and TX$_{1}$) from 4.5 K to 40 K was also performed on an FZ silicon sample which was processed using fabrication steps i through v. The $\mathrm{\upmu}$PL spectra are shown in the Supplementary Information Section C. 

Figure \ref{fig:fig3}b presents the temperature dependencies of the $^*$Cu$_0^0$, $^*$Cu$_0^1$, and T center ZPL transition energies that were extracted from Gaussian fits of their spectra. At cryogenic temperatures, the thermally induced redshift of the semiconductor bandgap can be expressed by the four parameter model,
\begin{equation}
    E(T)=E(0)-\alpha\Theta/2p\ (2T/\Theta)^p 
\end{equation}
where $\alpha$ denotes the slope $\sfrac{dE}{dT}$ in the limiting case of $T\rightarrow\infty$, $p$ is a dimensionless parameter related to the phonon dispersion coefficient $\sqrt{\frac{\langle\epsilon_{ph}^2\rangle-\langle\epsilon_{ph} \rangle^2}{\langle\epsilon_{ph} \rangle}}$, and $\Theta$ approximately equals the average phonon temperature \cite{passler1999parameter}. For simple fitting purposes, Eq. (1) can be rewritten as a polynomial expression,

\begin{equation}
    E\left(T\right)=E\left(0\right)-AT^p.
\end{equation}

The transition energies of $^*$Cu$_0^0$, $^*$Cu$_1^0$, and T centers were fitted using Eq. (2) and represented by the black curves in Figure \ref{fig:fig3}b. The dimensionless fitting parameter $p$ is equal to 1.25, 1.23, 2.27, and 2.36 for $^*$Cu$_1^0$, $^*$Cu$_0^0$, TX$_{1}$, and TX$_{0}$ transitions respectively. For comparison purposes, since the defect centers were in a sea of silicon atoms, we assumed that the average phonon temperature was the same and equaled $\Theta\approx 406$ K  for silicon \cite{passler1999parameter}. We then extracted the value of $\alpha$ to be 31.2 $\mathrm{\upmu}$eV/K, 16.8 $\mathrm{\upmu}$eV/K, 262 $\mathrm{\upmu}$eV/K, and 295 $\mathrm{\upmu}$eV/K for $^*$Cu$_1^0$, $^*$Cu$_0^0$, TX$_{1}$, and TX$_{0}$ respectively. The silicon band gap variation with temperature, plotted as grey dashed lines along with the $^*$Cu$_1^0$, $^*$Cu$_0^0$, and T center transition energies, was calculated by using the following parameters for silicon: $p$ = 2.33, $\alpha=318$ $\mathrm{\upmu}$eV/K, and $\Theta\approx406$ K\cite{passler1999parameter}. TX$_{0}$ and TX$_{1}$ parameters were close to silicon values, indicating a smaller modification of the silicon matrix by the T center in the sample where steps i through v were performed. At the end of our processing steps, our analysis suggested that the lattice was effectively restored to the pre-processing state.

In stark contrast, the red shifting behavior of $^*$Cu$_0^0$ and $^*$Cu$_1^0$ transition energies exhibited a dramatic deviation from the silicon bandgap, resulting in a much smaller fitting parameter of $p$ and $\alpha$ compared to the silicon and T centers. This further confirmed that a significant distortion of the silicon matrix occurred during step i, where the injected carbon ions and displaced silicon atoms occupied the interstitial sites, creating vacancies in the implantation path. Step ii promoted the impurity migration and resulted in the formation of 0.56 meV-split doublet luminescence centers. A much lower value of $p$ and $\alpha$ resulted in a more significant phonon dispersion. In Figure \ref{fig:fig3}c, we observed that the energy difference between the $^*$Cu$_1^0$ and $^*$Cu$_0^0$ lines was reduced by at least 0.1 meV until approximately 30 K, before quenching. The significant phonon dispersion led to a higher dependence of doublet splitting of the $^*$Cu$_0^0$ and $^*$Cu$_1^0$ lines with respect to temperature as compared to the T center doublet.  On the other hand, the reduction in energy difference between TX$_{0}$ and TX$_{1}$ was less dependent upon increasing temperature, owing to the minor modification of the silicon matrix.

Figure \ref{fig:fig3}d shows the integrated PL intensities of the $^*$Cu$_0^0$, $^*$Cu$_1^0$, and T center ZPLs with increasing temperature. TX$_{0}$ and TX$_{1}$ began to quench around 20 K, while the $^*$Cu$_0^0$ line quenched much earlier. The intensity of the $^*$Cu$_1^0$ line exhibited a plateau up to 15 K. The measured $1.75\pm0.01$ meV spectroscopic splitting between the TX$_{0}$ and TX$_{1}$ lines at 4.5 K matched the previously reported value \cite{irion1985defect}. The reported disassociation energy of the bound exciton was 22.5 meV \cite{irion1985defect}. The TX$_{0}$ intensity (black dots in Figure \ref{fig:fig3}e) was fitted using the following thermal partition function,
\begin{equation}
I(T)=\frac{I(0)}{1+Aexp(-\frac{E_1}{kT})+BT^{3/2}exp(-\frac{E_2}{kT})}
\end{equation}
by setting $E_1=1.75$ meV, as the excess energy of the excited state TX$_{1}$ to TX$_{0}$, and $E_2 = 22.5$ meV, as the activation energy to the silicon band edge. This is shown as the black curve in the upper panel of Figure \ref{fig:fig3}g. The integrated PL intensity ratio $I_{^*Cu_1^0}/I_{^*Cu_0^0}$ (inset of Figure \ref{fig:fig3}a) with respect to inverse temperature, yielded a thermal activation energy ($E_a$) of 0.44 $\pm{}$ 0.12 meV. This agreed with our measured spectroscopic splitting value of 0.56 $\pm{}$ 0.01 meV that was within our data fitting error. The aforementioned observation suggests that, like TX$_{1}$ and TX$_{0}$, the $^*$Cu$_1^0$ line can be associated with a transition from the higher excited state to the lower $^*$Cu$_0^0$ state of the same defect center. As shown in the lower panel of Figure \ref{fig:fig3}e, Eq. (3) can be used to fit the $^*$Cu$_0^0$ intensity by setting $E_1$ to be the measured spectroscopic doublet splitting of 0.56 meV, yielding an $E_2$ value of  $3.35$ meV after the fit converged. This suggests a much smaller binding energy than the T center bound excitons and agrees with the much earlier intensity quenching of the $^*$Cu$_0^0$ line even below 10 K. The total electron and hole binding energy $E_B$, associated with the bound exciton localized to this defect, was calculated to be 225.2 meV by subtracting the transition energy of $^*$Cu$_0^0$ by the silicon bandgap at low temperature. Like the T center, $E_B\gg E_2$ probably suggests that this 0.56 meV-split doublet related luminescence center consists of one strongly localized and another loosely Coulombic bound particle. This is potentially another candidate for a long lifetime spin-photon interface in the telecommunication band \cite{hautierPRM2022}. Note that the $\mathrm{\upmu}$PL intensity plateau of the $^*$Cu$_1^0$ line below 15 K was likely due to the thermal induced carriers from the lower energy $^*$Cu$_0^0$ state populating the higher energy $^*$Cu$_1^0$ state.

Figure \ref{fig:fig3}f shows the temperature dependent variation of FWHM linewidths. Both the lower energy states, TX$_{0}$ and $^*$Cu$_0^0$ exhibited constant linewidths of approximately 542 $\mathrm{\upmu}$eV and 318 $\mathrm{\upmu}$eV, respectively, within experimental error. However, for the higher energy states, the TX$_{1}$ linewidth stayed constant at approximately 598 $\mathrm{\upmu}$eV until around 20 K and then increased to 927 $\mathrm{\upmu}$eV at 40 K. The $^*$Cu$_1^0$ linewidth increased from 263 $\mathrm{\upmu}$eV to 444 $\mathrm{\upmu}$eV from 4.5 K to 15 K. It then gradually decreased to 343 $\mathrm{\upmu}$eV at 30 K. The observed broadening with respect to the Fourier transform limited linewidth of T center ($\leq$ 1 neV) could have resulted from: (1) nearby fluctuating charges which could have led to pure exciton dephasing or spectral diffusion, depending upon the rate of energy Stark shift, (2) ensemble emitting centers where individual ZPLs were shifted due to different local strains, (3) isotopic effects from natural silicon, in which the $^{29}$Si with non-zero nuclear spin caused resonance shift and magnetic field fluctuations, and (4) phonon-assisted broadening. However, for temperature variation of linewidths in general, phonon contribution is not significant for T = 30 K. We thus attributed the broadening of TX$_{1}$ with temperatures $\leq$ 30 K, to the increasing charge fluctuations associated with thermally induced electrons resulting from increasing temperatures. The minor variation of the TX$_{0}$ linewidth compared to the TX$_{1}$ linewidth can be due to its more localized exciton wavefunction. This also applies to the constant behavior of linewidth for the lower energy $^*$Cu$_0^0$ state compared to the $^*$Cu$_1^0$ state. The $^*$Cu$_1^0$ linewidth gradually increased up to 15 K, then decreased and finally quenched above 30 K as shown in Figure \ref{fig:fig3}f. This whole behavior could be attributed to increased charge fluctuations from the thermally induced carriers, considering the small activation energy of 0.56 meV for charge transfer from $^*$Cu$_0^0$ to $^*$Cu$_1^0$ state. The decrease in the $^*$Cu$_1^0$ linewidth can be due to charge stabilization of the fully occupied $^*$Cu$_1^0$ states at higher temperatures.

\subsection{Photoluminescence dynamics of T center and $^*$Cu defects}

\begin{figure*} \includegraphics[width=0.75\textwidth]{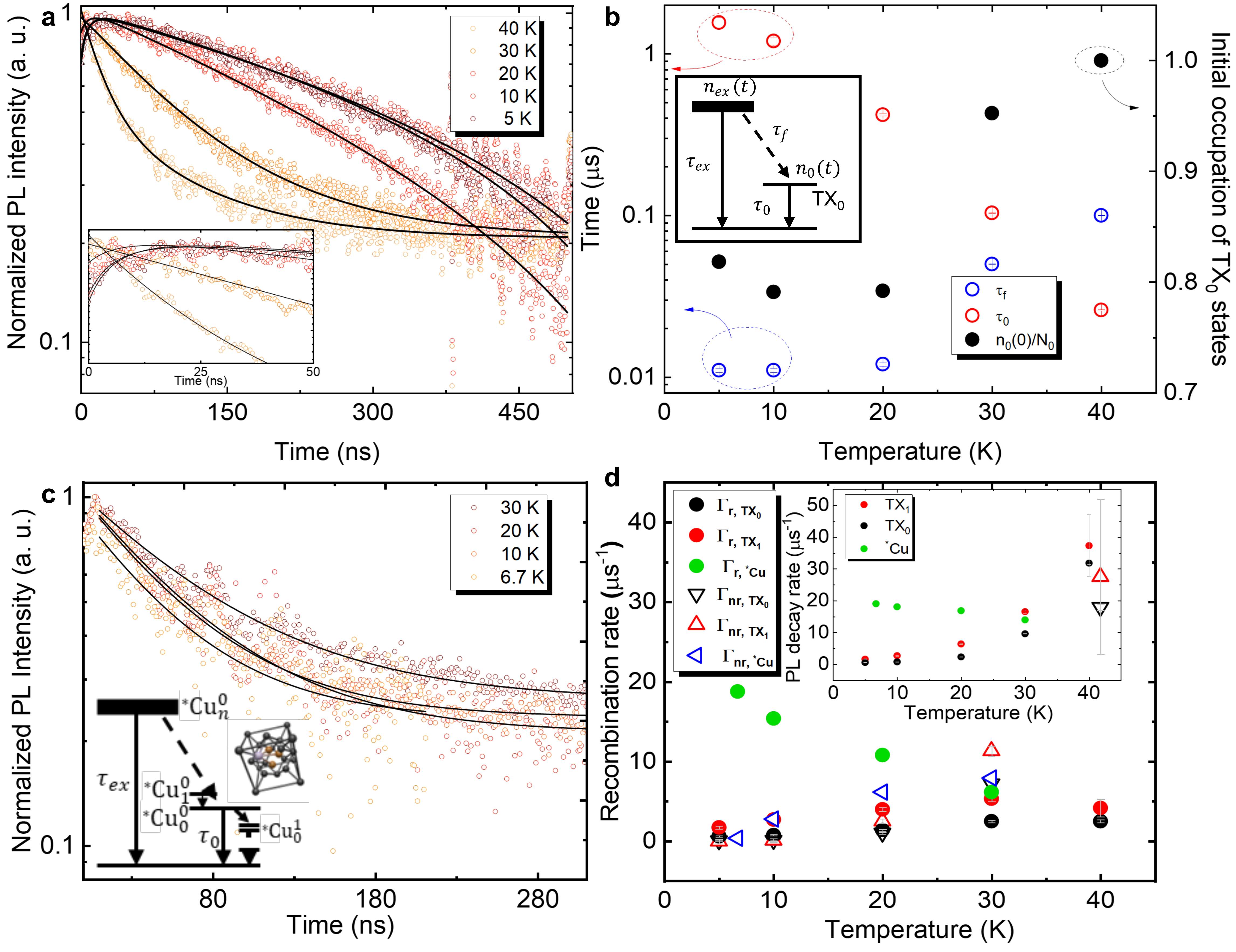}
    \caption{\textbf{T center and  $^*$Cu$^{m}_{n}$ bound exciton radiative dynamics.} \textbf{(a)} Time-resolved photoluminescence (TRPL) trace of TX$_0$ as a function of temperature fitted rate equation. Inset: zoom-in of early TRPL rise time. \textbf{(b)} Radiative lifetime and carrier populations as a function of temperature, with the TX$_0$ state recombination mechanism depicted in the inset. The carrier populations of TX$_0$ and higher excited states are denoted as $n_0(t)$, and $n_{ex}(t)$  with corresponding lifetimes of $\tau_0$ (red circles, left y-axis) and $\tau_{ex}$. $\tau_f$ denotes the relaxation time constant from the higher excited state to TX$_{0}$ (blue circles, left y-axis), and $N_0$ denotes the maximum available TX$_0$ states (black dots, right y-axis). \textbf{(c)} TRPL trace of the combined doublet decay of $^*$Cu$^{m}_{n}$ exciton. \textbf{(d)} Bound exciton recombination rate as a function of temperature. $\Gamma_{nr,TX_0}$ (black triangles), $\Gamma_{nr,TX_1}$ (red triangles), $\Gamma_{nr,^*Cu}$ (blue triangles) ,$\Gamma_{r,TX_0}$ (black dots), $\Gamma_{r,TX_1}$ (red dots),  $\Gamma_{r,,^*Cu}$ (green dots) denote the nonradiative and radiative recombination rate of TX$_0$,  TX$_1$,
    and $^*$Cu$^{m}_{n}$ accordingly.  Inset: The inverse PL decay time of TX$_0$(black dots), TX$_1$ (red dots), and $^*$Cu$^{m}_{n}$ (green dots) as a function of temperature.  The error bars in the figure denote the standard deviation of the fitting parameter.}
    \label{fig:fig4}
\end{figure*}

\textcolor{black}{ The recombination lifetime for TX$_{0}$ was previously measured to be 0.94 $\mathrm{\upmu}$s \cite{simmons2020}, while its calculated recombination lifetime was between 2 $\mathrm{\upmu}$s to 10 $\mathrm{\upmu}$s \cite{hautierPRM2022}. Accordingly, we performed temperature dependent time resolved photoluminescence (TRPL) measurements on TX$_{0}$ and TX$_{1}$ color center transitions to elucidate their radiative and nonradiative bound exciton dynamics.} We excited samples with a pulsed laser (600 nm, pulse width $\approx$ 5 ps) at an angle of approximately 45$\degree$ to the sample surface via a plano-convex singlet lens, creating a large excitation spot diameter of approximately 40 $\mathrm{\upmu}$m. We achieved sufficient counts for the TRPL histogram integration while refraining from saturating the T centers. The $\mathrm{\upmu}$PL was collected from the excitation spot via a 20$\times$ (NA = 0.4) microscope objective perpendicular to the sample surface and passed through a 0.5 m long monochromator to a near-infrared (NIR) photomultiplier tube (PMT) unit with $\approx$ 400 ps timing resolution attached to the exit port. The TRPL was then performed using the time correlated single photon counting (TCSPC) technique. The high-density 1200 grooves/mm grating used achieved a spectral resolution of 0.04 nm so that only the ZPL of TX$_{0}$ or TX$_{1}$ fed into the PMT, and the phonon sideband photons were rejected.

Figure \ref{fig:fig4}a shows the resulting TRPL traces of TX$_{0}$ transition with sample temperatures from 5 K to 40 K. At low temperatures, the TX$_{0}$ $\mathrm{\upmu}$PL decay exhibited a prolonged tail and the trend remained similar with increasing temperature until reaching 20 K, after which it changed to a faster decay. As shown in the inset of Figure \ref{fig:fig4}a, a slow rise of the TX$_{0}$ PL intensity was visible at 5 K and 10 K and gradually vanished with increasing temperature. \textcolor{black}{ We speculate that the carrier transition to TX$_{0}$ from the higher excited states (TX$_{2}$ to TX$_{25}$) led to a slow rise of the photoluminescence to its peak \cite{simmons2020}. We phenomenologically explain this process using a simple two level system coupled to the continuum of states, as shown in the inset of Figure \ref{fig:fig4}b. Such a slow rise component of the $\mathrm{\upmu}$PL intensity and its variation with temperature was not observed in TX$_{1}$ states nor G centers (see Supplementary Information Section E and F), nor reported in prior W \cite{dreauACS2022}, G \cite{CassaboisNatElec2020,CassaboisPRB2018} and T center studies \cite{simmons2020,simmonsArxiv2021,simmonsNJP2021,simmons2020,cavity1,cavity2}. The lack of observation by the earlier works\cite{simmons2020} for TX$_{0}$ could be due to a combination of the following factors: (1) the near silicon band gap laser used for excitation ($\lambda$ = 965 nm) \cite{simmons2020}, which led to fewer photogenerated carriers, (2) the temperature during TRPL, which was around 1.2 K, leading to lower population at higher energy levels, and (3) the 40 ns time resolution limitation in the measurement, which was not sufficient to capture the tens of nanosecond dynamics of a slow rise in the TRPL trace \cite{simmons2020}.}

\textcolor{black}{We rule out the possibility of an experimental artifact in our study for the following reasons: (1) we eliminated the ambient light contamination and phonon sideband contribution using a combination of 1325 nm, 50 nm bandwidth bandpass filter, and a 1200 grooves/mm grating with a spectral resolution of 80 $\mathrm{\upmu}$eV to isolate the zero-phonon line. (2) We observed temperature-dependent changes in TX$_{0}$ decay, which is shown and analyzed in Figures \ref{fig:fig4}a,b and d, and explained further in following paragraphs. And, (3) using the same experimental apparatus, we did not observe such slow \textcolor{black}{rise} in temperature-dependent TRPL measurements of $^*$Cu and G-centers.}

The measured $\mathrm{\upmu}$PL decay of TX$_{0}$ can be interpreted as the carrier relaxation from TX$_{0}$ to a ground state with a lifetime $\tau_0$ (radiative and nonradiative) and a nonradiative carrier feeding from an effective higher excited state $n_{ex}(t)$ to TX$_{0}$ with a characteristic time constant $\tau_f$. Note that since the carrier transition from the silicon band edge was fast (tens of fs), \cite{sjodin1998ultrafast,worle2021ultrafast} it could not be resolved in the measurement and thus was absorbed into the initial carrier populations $n_0(0)$ and $n_{ex}(0)$.
The following rate equations describe the carrier populations of $n_0(t)$ and $n_{ex}(t)$:
\begin{align}
   	\frac{dn_0}{dt}=-\frac{n_0}{\tau_0}+\frac{n_{ex}}{\tau_f}\left(\frac{N_0-n_0}{N_0}\right) \\ \frac{dn_{ex}}{dt}=-\frac{n_{ex}}{\tau_{ex}}-\frac{n_{ex}}{\tau_f}\left(\frac{N_0-n_0}{N_0}\right) 
\end{align}
where $n_0$ and $n_{ex}$ are populations of the optically excited carriers in TX$_{0}$ and higher excited states respectively. $\tau_0$ is the recombination lifetime of the TX$_{0}$ state, and $\tau_{ex}$ is the effective lifetime of higher excitation states. $\tau_f$ is the characteristic time constant of the nonradiative carrier feeding to TX$_{0}$ states. $N_0$ denotes the total amount of available TX$_{0}$ states that can be occupied.

As shown in Figure \ref{fig:fig4}a, the TRPL trace of TX$_{0}$ was adequately fit by equations, and the extracted fitting parameters are shown in Figure \ref{fig:fig4}b as a function of temperature. We extracted a recombination lifetime of $\tau_0=1.56 \pm{} 0.13$ $\mathrm{\upmu}$s at 5 K. With increasing temperatures, $\tau_0$ decreased to $26.0 \pm{} 0.2$ ns at 40 K, where TX$_{0}$ emission almost vanished. Importantly, we observed the carrier transition lifetime of $\tau_f=11.0 \pm{} 0.3$ ns at 5 K and 10 K. This low value of $\tau_f$ contributed to the demonstrably slow rise of TRPL traces. The observed 11 ns carrier transition and the resulting slow $\mathrm{\upmu}$PL rise (only observed for TX$_{0}$ below 10 K) was caused by a much more localized nature of TX$_{0}$ states. $\tau_f$ was found to be three orders of magnitude faster than the recombination lifetime of TX$_{0}$, indicating a strong coupling between the localized TX$_{0}$ and the delocalized higher excited states. At a temperature of 30 K the value $\tau_f$ was found to have increased to $50.0 \pm{} 0.3$ ns.

Considering that the pump power was fixed for TRPL measurement at all temperatures, the ratio of initial occupation of TX$_{0}$ states to its total amount of available states, $\frac{n_0(0)}{N_0}$, showed a sharper increase at 30 K, indicating the start of thermal disassociation of TX$_{0}$ bound exciton. Note that the timing resolution of the detector ($\approx$ 400 ps) used was much faster than the lifetimes of the carriers, and hence the detector's resolution can be considered negligible in the fit. We also performed the rate equation fitting by setting $n_{ex}=n_1$, where $n_1$ denotes the TX$_{1}$ carrier population, but the simultaneous fitting of TX$_{1}$and TX$_{0}$ $\mathrm{\upmu}$PL decay trace could not be achieved. This agrees with the hypothesis that the nonradiative carrier transition is dominated by higher excited states, which also corresponds to our case of the above silicon bandgap pumping. The coupling between TX$_{1}$ and TX$_{0}$ states can be a much faster process that can only be investigated by resonantly pumping the TX$_{1}$ state. 

Figure \ref{fig:fig4}c shows the temperature dependent TRPL of $^*$Cu$_0^0$-$^*$Cu$_1^0$ lines. Due to the small 0.56 meV-splitting and experimental limitations, both $^*$Cu$_0^0$ and $^*$Cu$_1^0$ lines were measured together with a TRPL window between 1309 nm and 1316 nm. With the presence of higher distortion in the lattice in Figure \ref{fig:fig3}b (increased deviation of dashed model lines from the temperature dependent fit of measured $^*$Cu$_n^m $data from Si lattice), the $^*$Cu$_0^0$-$^*$Cu$_1^0$ recombination occurs at a faster rate, with a lifetime $\tau_0$ of $52 \pm{} 5$ ns at 6.7 K. We observed an increase in the fitted lifetime to $71 \pm{} 2$ ns at 30 K, due to the increased population transfer between $^*$Cu$_0^0$ and $^*$Cu$_1^0$ and other higher excited states with slower recombination rates. To further investigate the T center bound exciton recombination dynamics, we assumed that both the radiative and nonradiative recombination channels contributed to the $\mathrm{\upmu}$PL recombination rate. They are specified by their respective recombination rates $\Gamma_r$ and $\Gamma_{nr}$.The time integrated $\mathrm{\upmu}$PL intensity $I$ of TX$_{0}$ and TX$_{1}$ at temperature T is proportional to the quantum efficiency 
$I(T)\propto\frac{\Gamma_r}{\Gamma_r+\Gamma_{nr}}$. With the radiative recombination dominated by the bound exciton recombination channel at T = 0 K,  $I(T)$ can be written as:
\begin{equation} I\left(T\right)=I\left(0\right)\frac{\Gamma_r}{\Gamma_r+\Gamma_{nr}}=I\left(0\right)\frac{\Gamma_r}{\Gamma_{tot}}.
\end{equation}

In the equation, $I(0)$ was extrapolated from data in Figure \ref{fig:fig3}d. $\Gamma_{tot}$ is the recombination rate extracted from the rate equation fit of TX$_{0}$ or the single exponential fit of TX$_{1}$. Then, the radiative recombination rate was obtained from equation (6). The resulting $\mathrm{\upmu}$PL dynamic parameters of TX$_{0}$ and TX$_{1}$ as a function of temperature were plotted in Figure 4d. The recombination rate of TX$_{1}$ ($\tau_1=0.60\pm0.06$ $\mathrm{\upmu}$s) was nearly twice of TX$_{0}$ at 5 K and gradually increased up to 30 K in line with TX$_{0}$ behavior. The extracted radiative and nonradiative bound exciton recombination rates of TX$_{0}$ and TX$_{1}$, with the experimental error bars, are shown in the inset. Compared with the nonradiative rate, the radiative rates of both TX$_{0}$ (black circle) and TX$_{1}$ (red circle) were found to be relatively constant $\approx$ 1.5 $\mathrm{\upmu}$s$^{-1}$ and 3.6 $\mathrm{\upmu}$s$^{-1}$, respectively, corresponding to a Fourier transform limited linewidth of $\approx$ 0.494 neV and 1.19 neV. 

The relatively temperature-independent rates (below 30 K) indicate potentially the suppression of thermally induced fluctuation of the bound exciton wavefunction along any direction due to the three-dimensional quantum confinement\cite{rosalesPRB2013}. A higher excited state (TX$_{1}$) is theorized to be more delocalized, characterized by a larger TX$_{1}$ radiative rate. The slight increase in both TX$_{0}$ and TX$_{1}$ radiative rates with increasing temperature was due to the thermally induced delocalization of the bound exciton wavefunctions. Above 30 K, the fast nonradiative recombination was observed to dominate both TX$_{0}$ and TX$_{1}$ channels. This could be due to the thermal disassociation of bound excitons, or the Shockley-Read-Hall (SRH) process, induced by unwanted impurities or vacancies in natural Si, which results in the quenching of the T center $\mathrm{\upmu}$PL intensity above 30 K. 

On the other hand, the radiative recombination rate of combined $^*$Cu$^{m}_{n}$ emissions showed abnormal behavior, reducing with increased temperature from 18.76 $\mathrm{\upmu} $s$^{-1}$ at 6.7 K to 6.13 $\mathrm{\upmu} $s$^{-1}$ at 30 K, while the nonradiative recombination rate increased to 7.88 $\mathrm{\upmu} $s$^{-1}$ and became the dominant decay channel. This counterintuitive phenomenon was attributed to the difference in decay rates and multiple phonon-assisted decay pathways (phonon replicas). We speculate that $^*$Cu$_0^0$ decay was faster than that of $^*$Cu$_1^0$, rendering the $^*$Cu$_0^0$ pathway dominant due to the increased population transfer at higher temperatures, as shown by the plateau in the $^*$Cu$_1^0$ plot observed in Figure \ref{fig:fig3}d. After 30 K, the radiative decay of both $^*$Cu$_1^0$ and $^*$Cu$_0^0$ is thermalized by nonradiative and phonon-assisted decay through $^*$Cu$_1^n$ states, which is outside of the spectral window of TRPL.

\subsection{Magnetic field induced broadening in the ensemble of $^*$Cu color centers}

\begin{figure*}[hbt!]    \includegraphics[width=0.75\textwidth]{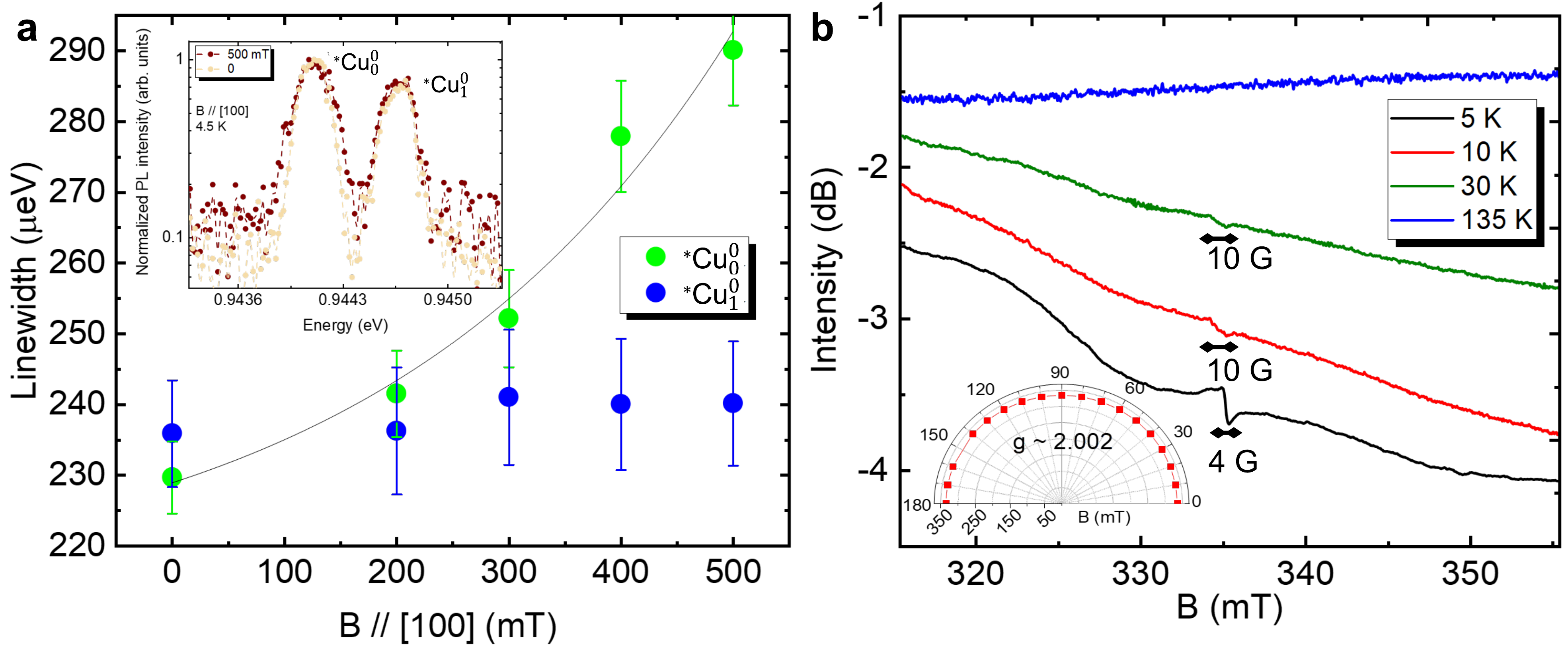}
    \caption{\textbf{Magnetic field induced broadening in the ensemble of 0.56 meV-split doublet luminescence centers.} \textbf{(a)} The Voigt profile linewidth of $^*$Cu$_0^0$ (green dots) and $^*$Cu$_1^0$ (blue dots) lines under a DC magnetic field with intensity up to 500 mT. Compared to the zero field linewidth of 230 $\mathrm{\upmu} $eV, we observed an increase in inhomogeneous linewidth of the $^*$Cu$_0^0$ line by 25\% at 500 mT, while there was not a significant broadening in the $^*$Cu$_1^0$ line. Inset: PL spectrum of $^*$Cu$_0^0$ and $^*$Cu$_1^0$ lines against DC magnetic field between 0 mT (yellow dots) and 500 mT (red dots). The linewidth of the $^*$Cu$_0^0$ line was fitted using a hyperbolic equation to determine the mean Zeeman factor as 364 $\mathrm{\upmu} eV/T$. The error bars in the figure denote the standard deviation of the fitted linewidth. Excitation laser: 532 nm, power: 900 W/cm$^2$. \textbf{(b)} Temperature dependent electron spin resonance measurements revealing a spin resonance at 335 mT, leading to an isotropic g-factor around 2.002. As predicted in the literature\cite{Lightowlers.1989}, the observations were consistent with an isotropic triplet state. Inset: Goniometric electron spin resonance spectrum to determine isotropy of the spin states.}
    \label{fig:fig5}
\end{figure*}

The T center bound exciton pairs with its ground state electron (which has a spin with an isotropic $g$-factor of $g_E=2.0005$) to form a singlet with an unpaired hole spin with an anisotropic $g$-factor in TX$_{0}$ state \cite{simmonsNJP2021}. The anisotropic hole spin leads to different $g$-factors and diamagnetic shift coefficients based upon the orientation of the defect in the lattice \cite{simmons2020}. Under a magnetic field, the unpaired hole spin undergoes Zeeman splitting and diamagnetic shifts. The total shift in the peak energy for spin dependent lines can be identified as $\Delta_E=(\pm g_e\pm g_h)\mathrm{\mu}_B B_z+\chi |B_z|^2$, where $g_e$ and $g_h$ are the $g$-factors for electrons and holes, $\mathrm{\upmu}_B$ is the Bohr magneton, $\chi$ is the diamagnetic shift due to higher order Zeeman effect, and $B_z$ is the applied magnetic field \cite{mcguyer2015precise}. The splitting enables a spin-photon interface that is essential for quantum memory applications. For the purpose of exploring other spin-photon interfaces, we examined the 0.56 meV-split $^*$Cu doublet luminescence center under a magnetic field to assess their potential candidacy as a spin-photon interface.

For an ensemble of defects, each of the hole spins in a singlet state has a different $g$-factor due to the different orientations in the lattice. This leads to an additional inhomogeneous broadening of the PL spectrum under the magnetic field alongside lifetime broadening, thermal broadening, and spectral diffusion between individual defect emissions. The resultant amplitude $A$ of the spin-polarized photoluminescence has a Voigt-like profile that accounts for these effects, described by the equation below \cite{simmonsNJP2021,mcguyer2015precise}:

\begin{multline}
    A(B_z,\Delta_f) = \int_{-\infty}^{\infty}(\frac{1}{\sigma_G\sqrt{2\pi}}e^{-\Delta_f/2\sigma_G^2}) \\ \frac{\sigma_L/2\pi}{(\sigma_L/2)^2+(\Delta_f-\varepsilon_{eff}|B_z|)^2}d\Delta_f
\end{multline}

Here, $\Delta_f$ is detuning, $\sigma_G$ \& $\sigma_L $ are Gaussian-broadened and homogeneous linewidths, and $\varepsilon_{eff}$ is the effective average Zeeman factor dependent on the hole $g$-factor variations. This leads to an approximate Voigt linewidth $\Gamma=\sqrt{((\Gamma_{B=0})^2+(\varepsilon_{eff}|B_z|)^2)}$ \cite{simmonsNJP2021}.

To assess the possibility of a spin-photon interface, we measured the magnetic field induced broadening of the 0.56 meV-split doublet line using non-resonant photoluminescence spectroscopy under an out of plane DC magnetic field ranging from 0 mT to 500 mT at 4.5 K. Instrument limitations and the broadening induced by the natural Si substrate did not allow us to distinguish the spin degeneracy lifted transitions within the ensemble fully. Figure \ref{fig:fig5}a shows the magnetic field induced broadening of the ensemble $^*$Cu$_n^m$ color center transitions. The intrinsic Voigt linewidth at 4.5 K and 0 mT is measured as 229 $\mathrm{\upmu}$eV for the $^*$Cu$_0^0$ line and measured as 235.9 $\mathrm{\upmu}$eV for the $^*$Cu$_1^0$ line. While there was not a significant broadening observed for the $^*$Cu$_1^0$ line, a sizable linewidth increase was observed for the first excited state $^*$Cu$_0^0$, by 25\% to 290 $\mathrm{\upmu}$eV. This increase in the linewidth was much larger than the instrument resolution. Using Eq.7, we extracted the effective Zeeman factor $\varepsilon_{eff}$ as 364 $\mathrm{\upmu} $eV/T for the $^*Cu_0^0$ color center defect transition. 
Assuming that the $^*$Cu doublet line was similar to that of T centers, we expected the formation of an unpaired hole-spin, interacting with the ground state electron within the first excited transition of the $^*$Cu$_0^0$ line. Further electron spin resonance analysis, which is shown in Figure \ref{fig:fig5}b, revealed that the ensembles of $^*$Cu$_0^0$ and $^*$Cu$_1^0$ defects provided a lambda-type spin-photon interface, with a degenerate ground level split, an isotropic $g$-factor of 2.002, and a Lorentzian broadening of 4 G, consistent with earlier observation of the isotropic triplet nature of the $^*$Cu$_0^0$ peak \cite{Lightowlers.1989}. The ESR resonance signal at approximately 335 mT showed a Gaussian like broadening of 10 G above 10 K and disappeared completely above 30 K, consistent with the quenching of the $^*$Cu$_n^m$ PL lines. This further supports the preceding discussion that the magnetic field induced broadening in Figure \ref{fig:fig5}a is due to the spin degeneracy of the $^*$Cu line.

\section{Conclusions}
In this study, we have examined the process steps leading to the formation of $^*$Cu and T color centers. During the T center formation process, the well known G and W centers were already present after low dose carbon implantation. Subsequently, we observed that the G and W centers disappeared after the annealing process at 1000 \degree C, only to be replaced by a doublet peak at 1312.15 nm also referred to as the $^*$Cu$_0^0$ and $^*$Cu$_1^0$ peaks, both of which are isoelectronic bound exciton peaks similar to T centers and originating from a Cu-Ag transition metal color center. These optical transitions are present in the spectra of carbon implanted and annealed samples alongside perturbed Cu line phonon replicas. Furthermore, we have observed that the doublet $^*$Cu$_n^m$ optical transitions persist alongside TX$_{0}$ and TX$_{1}$ if H$_2$O treatment is omitted and only RTA is applied to the sample following hydrogen implantation. 

Furthermore, we have examined and compared the TX$_{0}$ and TX$_{1}$ with $^*$Cu$_0^0$ and $^*Cu_1^0$ transitions and investigated their origins. The temperature dependent $\mathrm{\upmu}$PL measurements verified that the $^*$Cu$_0^0$ and $^*$Cu$_1^0$ peaks resulted from lower and higher excited state transitions, respectively, of the same color centers whose formation induced a higher perturbation to the silicon host, compared to the less perturbative T centers. We further estimated a much smaller exciton binding energy associated with $^*$Cu defect, down to 3.35 meV.

Based on the abovementioned process, we examined the bound exciton dynamics of the generated $^*$Cu$_n^m$ and T color centers. With time resolved $\mathrm{\upmu}$PL measurements, we obtained radiative lifetimes of the  $^*$Cu$_n^m$ transition at 52 $\pm$ 5 ns, and for the T center at 1.56 $\pm$ 0.13 $\mathrm{\upmu} $s. We have confirmed our experimental results by fitting the rate equation modeling between TX$_{0}$ and higher excited states. We found a slow carrier transition between TX$_{1}$ and TX$_{0}$, corresponding to the initial period slow rising dynamics in our measured TRPL histograms. We further extracted the radiative and nonradiative recombination rates of the three-dimensional confined T center bound excitons. Reducing possible nonradiative decoherence pathways is of utmost importance for high fidelity quantum network nodes, while cavity quantum electrodynamics (QED) interactions can increase the coherent photonic qubit emission rates.

Lastly, we further studied the electron spin resonance (ESR) properties of the resolved $^*$Cu$_n^m $ doublet peak around 1312.15 nm, which shows a spin-degeneracy evidenced by the ESR spectroscopy, and the 25\% broadening observed under magnetic field. Thus, $^*$Cu represents a unique platform at the zero dispersion regime in the O-band for qubit interactions alongside T centers. The calculated excited state radiative lifetimes, and the magneto-optic and electron spin resonance measurements support the development of these silicon color centers for the purpose of realizing solid state quantum memories in scalable and metropolitan quantum networks\cite{redjemarxiv2023,Reisererarxiv2023}.

\section{Methods}
\textit{Power and temperature dependent microphotoluminescence spectroscopy |} A 532 nm solid state continuous wave (CW) laser was used for the above bandgap excitation with a continuous reflective neutral density (ND) filter for precise power control. A 925 nm dichroic mirror was used to reflect the incoming excitation laser to an objective (numerical aperture of 0.4) for focusing and PL collection. The laser spot on the sample was approximately 2 $\mathrm{\upmu}$m. The samples were housed in a Lakeshore Janis ST-500 microscopy cryostat with a quartz window vertically set up. The collected PL was passed through the dichroic mirror and then a 905 nm long pass filter for eliminating pump residual and the PL was directed into a Princeton Instruments 0.5 m long monochromator (SpectraPro 2500) with 600 grooves/mm grating. For magneto-PL measurement, DC permanent ring magnets with field intensities of 100 mT, 200 mT, 300 mT, 400 mT, and 500 mT were used to apply a vertical magnetic field perpendicular to the sample surface (parallel to [100]). The objective was installed on a precision optical rail, ensuring that the alignment was the same for each measurement after changing the applied field intensity.  A liquid nitrogen cooled InGaAs camera was used to measure individual lines with a spectral resolution of approximately 100 pm. The transmission efficiency for the following components were: objective lens $\approx$ 7\%, dichroic mirror $\approx$ 90\%, long pass filter $\approx$ 90\%, grating $\approx$ 55\%, and PMT coupling efficiency $\approx$ 90\%. Additionally, detection efficiencies for the following were: InGaAs array $\approx$ 80\% and PMT $\approx$ 2\%. Overall the collection efficiency for PL measurement was approximately 2.5\%. 

\textit{Time resolved microphotoluminescence spectroscopy |} The $\mathrm{\upmu}$PL spectroscopy setup was modified to utilize a pulse laser at 600 nm selected from a supercontinuum source (NKT Photonics SuperK Extreme) at a 2 MHz repetition rate, which created a 40 $\mathrm{\upmu}$m spot on the sample with an angle of approximately 45$\degree$ for side pumping. The focal points of the side excitation path and the collecting Mitutoyo objective were aligned by co-propagation from both beam paths. The collected light was passed through the same spectrometer with 1200 grooves/mm grating (spectral resolution $\approx$ 40 pm) and directed to a photomultiplier tube (PMT). The RF synchronized signal and PMT output were connected to a Swabian Instruments (TimeTagger 20) time-to-digital converter, achieving an overall timing resolution of 400 ps. The collection efficiency for TRPL measurements was approximately 0.06\%.

\textit{Cryogenic ESR measurement |} The cryogenic ESR measurement of Cu related defects was performed using Bruker EMXplus EPR spectrometer. The microwave probe frequency was set at 9.381 GHz, and the magnetic field was swept between 40 mT to 600 mT.

\section{Data Availability} The data that support the findings of this study are available from the authors upon reasonable request. 

\section{Acknowledgements} The authors thank Joshua Pomeroy, Louis Bouchard, Madeline Taylor, Jasmine Mah, Lloyd Mah, and Kerry Kangdi Yu for beneficial discussions regarding the paper. The authors acknowledge support from the Army Research Office Multidisciplinary University Research Initiative (W911NF-21-2-0214), the National Science Foundation under award numbers 2141064 (GRFP), 2125924 (NRT), 2137984 (QuIC-TAQS) and 1936375 (QII-TAQS). This work was performed, in part, at the Center for Integrated Nanotechnologies, an Office of Science User Facility operated for the U.S. Department of Energy (DOE) Office of Science. The authors acknowledge the usage of UC Santa Barbara MRL facilities for ESR spectroscopy measurements and thank Jaya Nolt for facilitating the experiments. The MRL Shared Experimental Facilities are supported by the MRSEC Program of the NSF under Award No. DMR 2308708 and is a member of the NSF-funded Materials Research Facilities Network.

\section{Author Contributions} M.C.S., J.H., and C.W.W. designed and led the project. M.C.S., C.F., and K.M.A-R. processed and fabricated the samples. M.C.S., J.H., J.H.K., C.F, K.M.A-R., and W. L. conducted photoluminescence spectroscopy experiments. M.C.S., J.H., C.F., K.M.A-R., and B.L. built the infrastructure and conducted time-resolved photoluminescence measurements. M.C.S., J.H.K, and J.H. conducted magnetic-field photoluminescence and electron-spin-resonance measurements. M.C.S., J.H., C.F., and K.M.A-R. contributed to data analysis. M.C.S, J.H., C.F., K.M.A-R., and C.W.W. wrote the manuscript, with inputs from all the authors.

\section{Competing Interests} The authors declare no competing interests.

\section{References}
\bibliography{bibliography}
\end{document}